\documentclass[english,APL,reprint,superscriptaddress]{revtex4-1}
\usepackage[latin9]{inputenc}
\usepackage{color}
\usepackage{textcomp}
\usepackage{amsmath}
\usepackage{amssymb}
\usepackage{graphicx}
\usepackage{subscript}
\usepackage{natbib}
\setcitestyle{square, comma, numbers,sort&compress, super}

\DeclareFontEncoding{LGR}{}{}

\ProvideTextCommand{\~}{LGR}[1]{\char126#1}

\usepackage{textcomp}

\usepackage{array}

\usepackage{babel}

\makeatother

\usepackage{babel}
\begin{document}
\title{Acceptor and compensating donor doping of single crystalline SnO(001) films grown by molecular beam epitaxy and its perspectives for optoelectronics and gas-sensing}
\author{Kingsley Egbo}
\affiliation{Paul-Drude-Institut f\"ur Festk\"orperelektronik, Leibniz-Institut im
Forschungsverbund Berlin e.V., Hausvogteiplatz 5--7, D--10117 Berlin,
Germany}
\author{Jonas L\"ahnemann}
\affiliation{Paul-Drude-Institut f\"ur Festk\"orperelektronik, Leibniz-Institut im
Forschungsverbund Berlin e.V., Hausvogteiplatz 5--7, D--10117 Berlin,
Germany}
\author{Andreas Falkenstein}
\affiliation{Institute of Physical Chemistry, RWTH Aachen, D-52056 Aachen, Germany}
\author{Joel Varley}
\affiliation{Lawrence Livermore National Laboratory, Livermore, California 94550, USA}
\author{Oliver Bierwagen}
\affiliation{Paul-Drude-Institut f\"ur Festk\"orperelektronik, Leibniz-Institut im
Forschungsverbund Berlin e.V., Hausvogteiplatz 5--7, D--10117 Berlin,
Germany}

\begin{abstract}
 (La and Ga)-doped tin monoxide (stannous oxide, tin (II) oxide, SnO) thin films were grown by plasma-assisted and suboxide molecular beam epitaxy with dopant concentrations ranging from $\approx5\times10^{18}$~cm$^{-3}$ to $2\times10^{21}$~cm$^{-3}$. In this concentration range, the incorporation of Ga into SnO was limited by the formation of secondary phases observed at $1.2\times10^{21}$~cm$^{-3}$ Ga, while the incorporation of La showed a lower solubility limit. Transport measurements on the doped samples reveal that Ga acts as an acceptor and La as a compensating donor. While Ga doping led to an increase of the hole concentration from $1\times10^{18}$~cm$^{-3}-1\times10^{19}$~cm$^{-3}$ for unintentionally (UID) SnO up to $5\times10^{19}$~cm$^{-3}$, La-concentrations well in excess of the UID acceptor concentration resulted in semi-insulating films without detectable $n$-type conductivity. Ab-initio calculations qualitatively agree with our dopant assignment of Ga and La, and further predict In$_\text{Sn}$ to act as an acceptor as well as Al$_\text{Sn}$ and B$_\text{Sn}$ as donor. These results show the possibilities of controlling the hole concentration in $p$-type SnO, which can be useful for a range of optoelectronic and gas-sensing applications.\\
 \\Email: \textit{\textcolor{blue}{egbo@pdi-berlin.de, bierwagen@pdi-berlin.de}}
\end{abstract}
\date{\today}
\maketitle

Reliable bipolar carrier transport remains a challenge in most transparent semiconducting oxides (TSOs), limiting the widespread adoption of oxides for optoelectronic devices.\cite{gkksssybh16} Many widely applied TSOs can readily be doped $n$-type, while their $p$-type doping remains challenging, if not untenable. However, few TSOs show $p$-type conductivity,\cite{eely20} with tin (II) oxide (SnO) being among them. Compared to other binary $p$-type TSOs, its optical bandgap of $\approx2.7$ eV and hole mobility of $\approx3$-5 cm$^2$/Vs, for phase-pure single crystalline (001) layers, makes it a candidate material for oxide complementary integrated circuits, $p$-channel thin film transistors, or transparent oxide heterojunctions with $n$-type materials.\cite{zxbr16,bsmtfrwgb20,tekupcagtbw22,pmbbesbstf22,Wang2015} Polycrystalline, Sn-rich SnO has even been shown to exhibit a room-temperature hole mobility of up to 30~cm$^2$/Vs.\cite{cnagsh13,mgamstg17} Conductometric gas sensors are another large application domain of TSOs. While SnO has not yet been widely explored for this application, $p$-type oxides are generally considered to allow for high sensitivity and selectivity.\cite{kim_highly_2014}

Unintentionally doped (UID) $p$-type conductivity in SnO is believed to originate primarily in Sn-vacancies\cite{tott06} or their complexes with hydrogen.\cite{vsjw13} Theoretical studies have also suggested the possibility for bipolar doping in SnO\cite{vsjw13,ggf18} and enhancement of $p$-type properties by native defects.\cite{gcas13} However, experimental studies exploring the effect of intentional impurities in SnO towards possible bipolar conductivity are still rare and partially conflicting:
Hayashi \textit{et. al.},\cite{hkhkt15} showed electron density of $3.0\times10^{15}$~cm$^{-3}$ in UID SnO.  Hosono \textit{et. al.}, showed that the doping of SnO with $>$8 cation at\% of Sb (N$_{Sb}$ ~$>2.4\times10^{21}$~cm$^{-3}$) can produce an electron density of ~$10^{17}$~cm$^{-3}$,\cite{hoyk11} whereas Guo \textit{et. al.}, demonstrated slight improvement in $p$-type properties by also Sb- 
doping.\cite{gfzzllcp10}

While acceptor doping of SnO should increase its hole concentration beyond the UID level, donor doping holds promises for obtaining $n$-type SnO conductivity or semi-insulating material which can be useful for desirable oxide $pn$ homojunctions or isolating buffer layers, respectively, for SnO-based devices.

Here, we revisit the possibility for reliable bipolar doping in single crystalline, phase-pure SnO thin films. An experimental study, backed by theoretical verification, on the effects of La and Ga dopants in SnO is presented. Transport measurements evidence that Ga acts as acceptors in SnO, which we attribute to a preferential incorporation of Ga$_{Sn}$ in SnO in the Ga$^{1+}$ oxidation state. La dopants, in contrast, show clear compensating donor behavior as highly La-doped SnO thin films became semi-insulating. These measurement results are corroborated by density functional theory calculations, which predict that In$_\text{Sn}$ and Ga$_\text{Sn}$ are soluble acceptors in SnO, while La$_\text{Sn}$ and other group III dopants like B$_\text{Sn}$ and Al$_\text{Sn}$ behave as donors.

Approximately 100-300 nm-thick UID and (La and Ga)-doped SnO (001) films were grown on YSZ(001) substrates by molecular beam epitaxy (MBE). We explored two growth techniques for SnO; plasma-assisted-MBE (PA-MBE) with a metallic Sn source and an activated oxygen source\cite{bmfgnuhhmrb20} and suboxide-MBE (S-MBE)\cite{ellhtgkfgb22} using a SnO$_2$+Sn mixture as efficient and pure SnO source\cite{hoffmann_efficient_2020} without the use of additional oxygen. In both growth techniques, the dopant elements were incorporated using the metal effusion cells and the dopant concentration was varied by changing the dopant cell temperatures for different growths. Hence, the dopant flux reaching the substrate, being proportional to the metal vapour pressure, is controlled by the metal cell temperature. All samples were grown at similar substrate temperatures between 350~$^{\circ}$C and 400~$^{\circ}$C. The grown films were analyzed in-situ by reflection high-energy electron diffraction (RHEED) and the thickness of the films was determined by in-situ laser reflectometry. The grown layers were structurally investigated ex-situ by x-ray diffraction (XRD) and atomic force microscopy (AFM). The dopant concentration was derived from energy dispersive x-ray spectroscopy (EDX) measurements and time of flight- secondary ion mass spectrometry (TOF-SIMS) as described in the \textcolor{blue}{Supplementary material}.  Detailed electrical properties of the doped films were obtained from van der Pauw-Hall measurements. Hybrid functional calculations of dopant formation energies with projector-augmented wave (PAW) potentials were performed using the HSE-screened hybrid functionals implemented in Vienna Ab Initio Simulation Package (VASP).\cite{hse06,kf96,kj99,b94} Defect calculations were performed using supercells adopting 192 atoms for the bulk SnO structure, with the same calculation parameters as described in Ref.\cite{vsjw13}. For the bulk and dopant PAW potentials, we consider Sn:[5s$^2$5p$^2$], In:[5s$^2$5p$^1$], Ga:[4s$^2$4p$^1$], Al:[3s$^2$3p$^1$], B:[2s$^2$2p$^1$], and La:[5s$^2$5p$^6$5d$^1$6s$^2$] electron configurations treated as valence states. Solubility limiting phases considered for the respective dopants were In$_2$O$_3$, Ga$_2$O$_3$, Al$_2$O$_3$, SnB$_4$O$_7$ and La$_2$Sn$_2$O$_7$.\\

Figure 1 shows the Ga- and La-concentrations as a function of dopant cell temperature for doped samples grown using PAMBE, as well as the Ga-concentration for Ga-doped samples from S-MBE growth. An increasing cell temperature coincides with increasing dopant concentration for all dopants. These dopant concentrations follow an Arrhenius-like behavior indicating efficient incorporation of the dopants in the film. While an activation energy of $\approx2.3$~eV is obtained for Ga-doped samples grown using S-MBE, i.e., without additionally supplied oxygen, the Ga activation energy obtained for samples grown using the PA-MBE method is almost two times lower. This points to a possibility that during doping with the elemental Ga source, the oxygen background during PA-MBE growth leads to the formation and  subsequent evaporation of the suboxide Ga$_2$O in the source, which has a lower activation energy in the range of 1.0-1.2 eV.\cite{hcbo21}

\begin{figure}
\includegraphics[width=8.5cm]{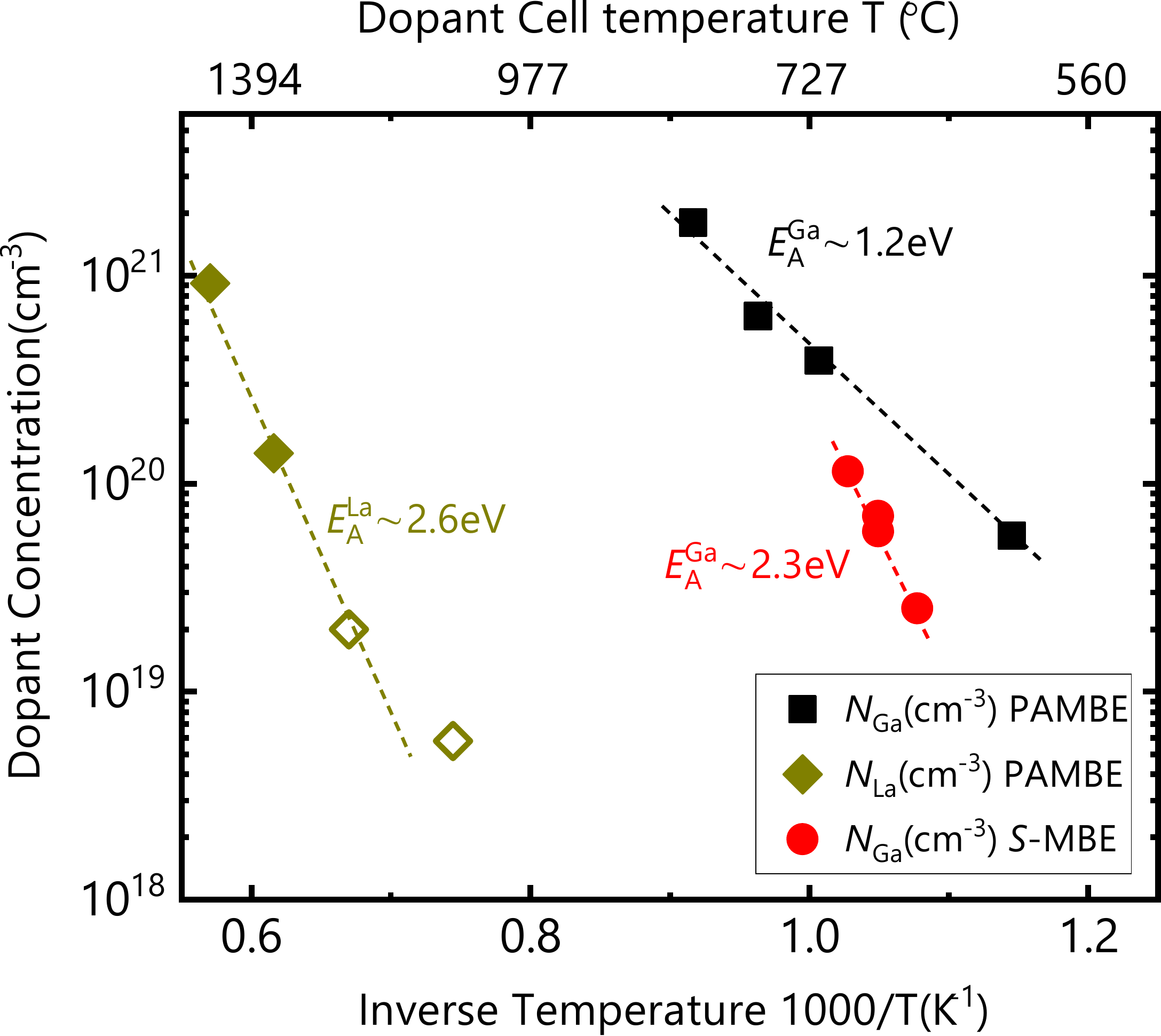}\centering
\caption{Dopant concentrations in the SnO films as a function of dopant cell temperature. Due to the detection limit of EDX, low La dopant concentrations with open symbol are obtained from TOF-SIMS and  growth rate calibration. The dotted lines indicate an Arrhenius-like behavior expected for the evaporation of these elements from their effusion cell. Extrapolated activation energies using these lines are also indicated. \label{fig:doping-levels}}
\end{figure}

\begin{figure*}
\includegraphics[width=14cm]{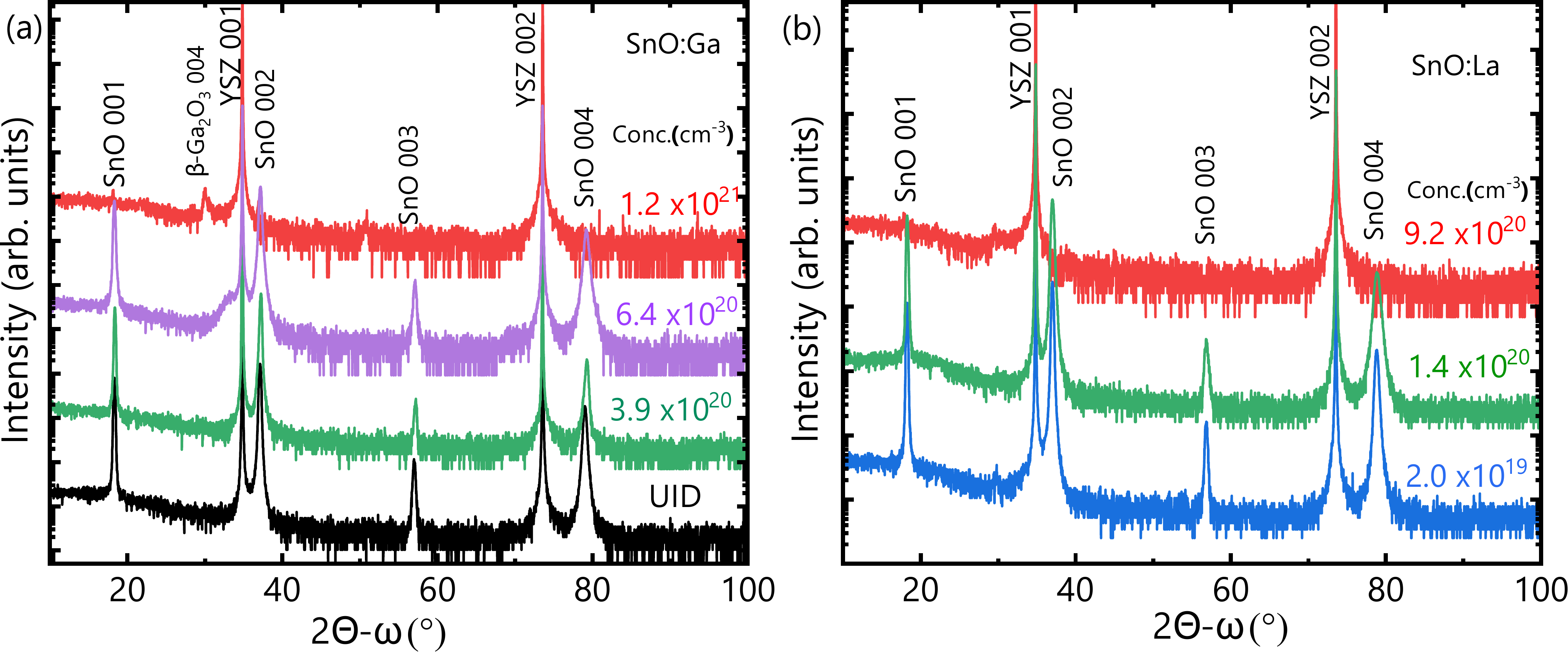}\centering
\caption{XRD 2$\theta$--$\omega$ scan of (a) UID and Ga-doped, and (b) La-doped SnO(001) thin films with different dopant concentrations grown by PA-MBE. \label{fig:xrd}}
\end{figure*}

The solubility limit of dopants in SnO was estimated from XRD. On-axis 2$\theta$-$\omega$ scans in Figure ~\ref{fig:xrd} show that phase-pure, single-crystalline SnO(001) films are obtained up to the solubility limit of the various dopants. XRD data in Figure 2(a) indicate phase-pure SnO(001) samples at Ga concentrations up to $6.4\times10^{20}$~cm$^{-3}$. At a higher Ga content of $1.2\times10^{21}$~cm$^{-3}$, a weak peak due to $\beta$-Ga$_2$O$_3$ (004) is present while the SnO related peak is absent. Similarly, in Figure 2(b), at a La concentration of $9.2\times10^{20}$~cm$^{-3}$, no SnO related peak is observed. Hence, in the concentration range studied, the incorporation of the extrinsic dopants was limited by the formation of secondary phases or amorphous layers. These XRD data indicate that the solubility limit of Ga and La in SnO are between $6.4 - 12\times10^{20}$~cm$^{-3}$ and $1.4-9.2\times10^{20}$~cm$^{-3}$, respectively. 
In Figure S1 (\textcolor{blue}{Supplementary material}), the AFM micrographs of an $\approx120$ nm UID layer grown at 350$^{\circ}$C by PA-MBE show fine and dense surface morphology films with a root-mean-square roughness of $\approx1.8$ nm. Approximately 300 nm thick, PA-MBE grown, Ga-doped layers also maintained similar morphology as the UID layer, however for La doping at a similar film thickness, slightly different morphology is observed which showed an increase in roughness and less coalesced grains.

\begin{figure*}
\includegraphics[width=15cm]{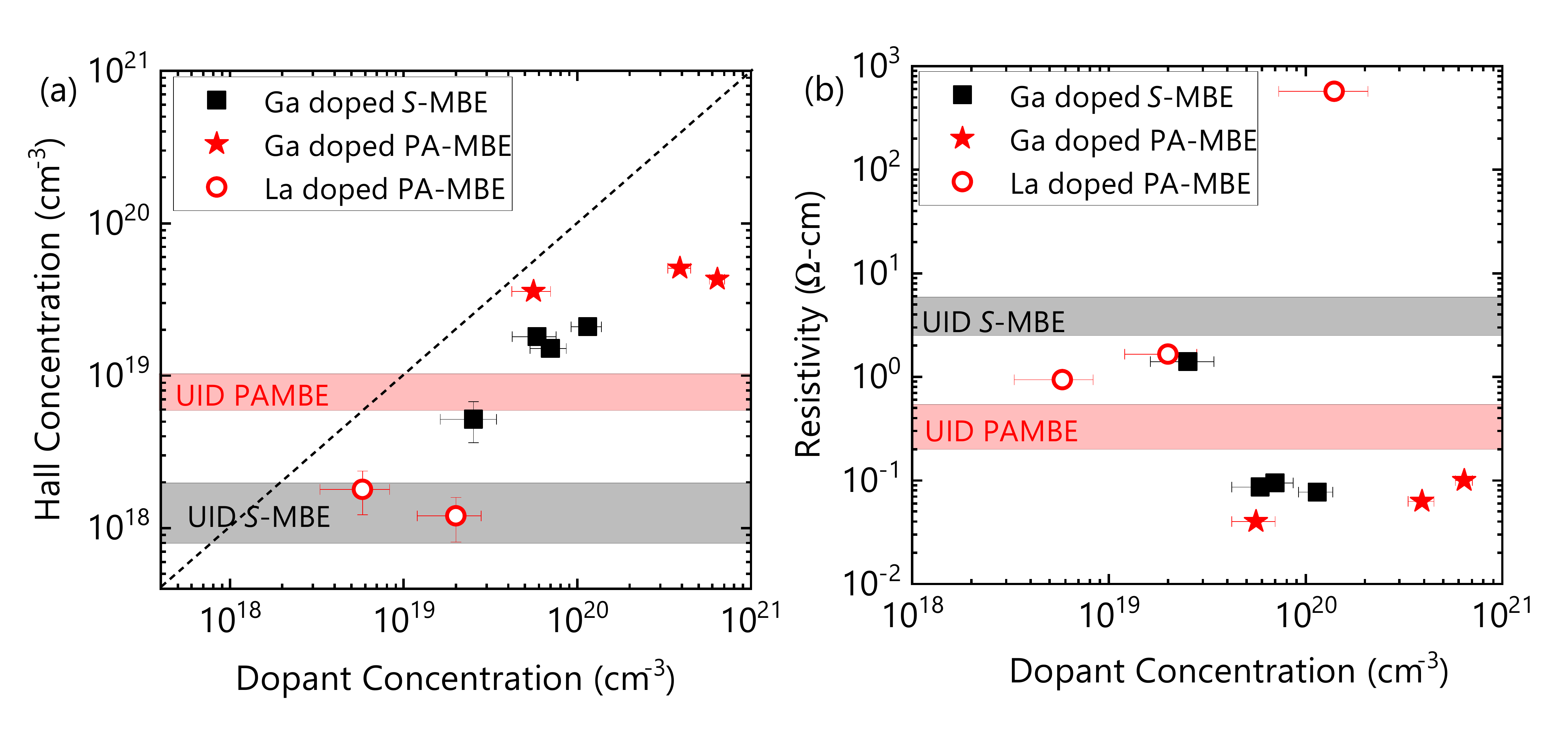}\centering
\caption{Hole concentration vs. dopant concentration (EDX) for Ga- and La-doped SnO thin films. The shaded regions indicate UID reference hole concentration values for PA-MBE (red) and S-MBE (black) samples. Ga-doped SnO thin films grown by S-MBE are also shown using black symbols, while all doped samples grown by PA-MBE are shown using red symbols. The dashed symbolizes a hole concentration equalling the dopant concentration which corresponds to an acceptor doping efficiency of unity and absent UID doping.  Semi-insulating, highly La-doped samples (La conc. $>1.0\times10^{20}$~cm$^{-3}$) could not be measured and data points are not shown. (b) Resistivity vs dopant concentration for PA-MBE (red) and S-MBE (black) samples. \label{fig:hall}}
\end{figure*}

Figure 3(a) and (b) show the Hall-measured hole concentrations and resistivities, respectively, of doped SnO thin films as function of the dopant concentration. For comparison, red and black shaded areas correspond to the range of properties of UID reference samples grown for this study via PA-MBE and S-MBE, respectively. The UID $p$-type conductivities obtained for SnO(001) layers obtained from S-MBE and PA-MBE are markedly different with higher hole concentrations for the PA-MBE grown films.
This difference in transport properties for the different growth methods could be due to the enhanced formation of Sn vacancies resulting from the energetics of the different growth processes with negligible dependence on the film growth rate. For S-MBE growth, SnO molecules formed in the crucible are reaching the substrate surface, which is expected to decrease the formation of Sn vacancies and complexes compared to the PA-MBE growth, where elemental Sn and activated oxygen are supplied to the substrate. Unintentional impurities in the MBE chamber may also play a role in the UID hole concentration as these values are higher than previously reported values for MBE-grown single crystalline SnO(001).\cite{gfzzllcp10,mmwkwbsnptzhsps19,bmfgnuhhmrb20}

The reference UID SnO(001) layers grown for this study under different conditions by PA-MBE exhibit room temperature (RT) Hall hole concentrations ($p_H$) in the range  of $\approx0.6$--$1.0\times10^{19}$~cm$^{-3}$, Hall hole mobilities ($\mu_H$) of $\approx$ 2.4-3.5 cm$^2$/Vs, and bulk resistivities ($\rho$) of $\approx0.73$--0.2 $\Omega$cm. We find that $p_H$ increases to $5.0\times10^{19}$~cm$^{-3}$  and $\rho$ decreases to 0.063 $\Omega$cm for  films doped with increasing amounts of Ga. SnO (001) films grown for this study by S-MBE show lower UID $p_H$ in the range of $0.62$--$2.0\times10^{18}$~cm$^{-3}$,\cite{ellhtgkfgb22} and $\rho=2.6$--5.0~$\Omega$cm, $p_H$ increases up to $3.0\times10^{19}$~cm$^{-3}$ and $\rho$ decreases to 0.08 $\Omega$cm with increasing Ga doping. In contrast, thin films doped with increasing concentrations of La show a reduction in $p_H$ and a remarkable increase of $\rho$ up to 580~$\Omega$cm  without transition to $n$-type conductivity. Samples with higher La doping became semi-insulating and were not measurable in our Hall setup.
\begin{table*}
\caption{\label{tab:table2}Electrical properties of unintentionally-doped ("UID") and corresponding intentionally-doped ("Doped") polycrystalline and single crystalline SnO thin films.}
\begin{ruledtabular}
\begin{tabular}{cccccccc}
 Material  &Grown by &UID~$p_H$(cm$^{-3}$) & Dopant Conc. (cm$^{-3}$)
 &Doped~$p_H$ (cm$^{-3}$) &Doped~$\rho$ ($\Omega$cm ) &Ref.\\
\hline
Ga-doped SnO(poly) & sputtering & $4.5\times10^{18}$ & Ga $\approx3.2\times10^{20}$ & $6.2\times10^{18}$ & 0.27 & \cite{Kwok2022} \\
Na-doped SnO(poly) & sputtering & $4.5\times10^{18}$ & Na $\approx8.6\times10^{20}$ & $1.1\times10^{19}$ & 0.12 & \cite{Kwok2022} \\
Y-doped SnO(001) & EBE\footnotemark[1] & $1.1\times10^{18}$ & Y $\approx1.4\times10^{21}$ & $2.7\times10^{18}$ & 3.8 & \cite{gfzzllcp10} \\
Sb-doped SnO(001) & EBE\footnotemark[1] & $1.1\times10^{18}$ & Sb $\approx1.4\times10^{21}$ & $1.3\times10^{18}$ & 5.5 & \cite{gfzzllcp10}  \\
Y-doped SnO(001) & EBE\footnotemark[1] & $5.6\times10^{15}$ & Y $\approx2\times10^{20}$ & $4.7\times10^{16}$ & 120 & \cite{llcxslk12} \\
Ag-doped SnO(poly) & sputtering & $1.0\times10^{18}$ & Ag $\approx1.4\times10^{21}$ & $1.3\times10^{19}$ & 0.5 & \cite{Pham2017} \\
Ga-doped SnO(001) & MBE\footnotemark[2] & $6.0\times10^{18}-1\times10^{19}$ & Ga = $4\times10^{20}$ & $5.0\times10^{19}$ & 0.063 & This work \\
La-doped SnO(001) & MBE\footnotemark[2] & $6.0\times10^{18}-1\times10^{19}$ & La = $1\times10^{20}$ & - & 580 & This work \\
\end{tabular}
\end{ruledtabular}
\footnotetext[1]{Electron Beam Evaporation}
\footnotetext[2]{Molecular Beam Epitaxy}
\end{table*}

Compared to previous studies on extrinsic doping in SnO using Ga, Na, Y, Sb, and Ag and their corresponding UID films, summarized in Table I, our Ga-doped SnO shows
the highest hole concentration.
Closer inspection of Figure 3(a), however, suggests that dopant efficiency in SnO is limited for the range of dopants in our study. The reduction in hole concentration is weaker than the increase of La (compensating donor) concentration. 
Likewise, the increase of hole concentration is well below that of the Ga (acceptor) concentration, and a strong saturation of the hole density is observed with increasing Ga concentration above $7.0\times10^{19}$~cm$^{-3}$ in both the S-MBE and PA-MBE growths.
Below a Ga concentration of $1.2\times10^{21}$~cm$^{-3}$, where Ga begins to form a secondary phase, transport data suggests that Ga dopants are not particularly shallow acceptors in SnO and likely not all acceptors can be ionized at RT. 

\begin{figure}
\includegraphics[width=8cm]{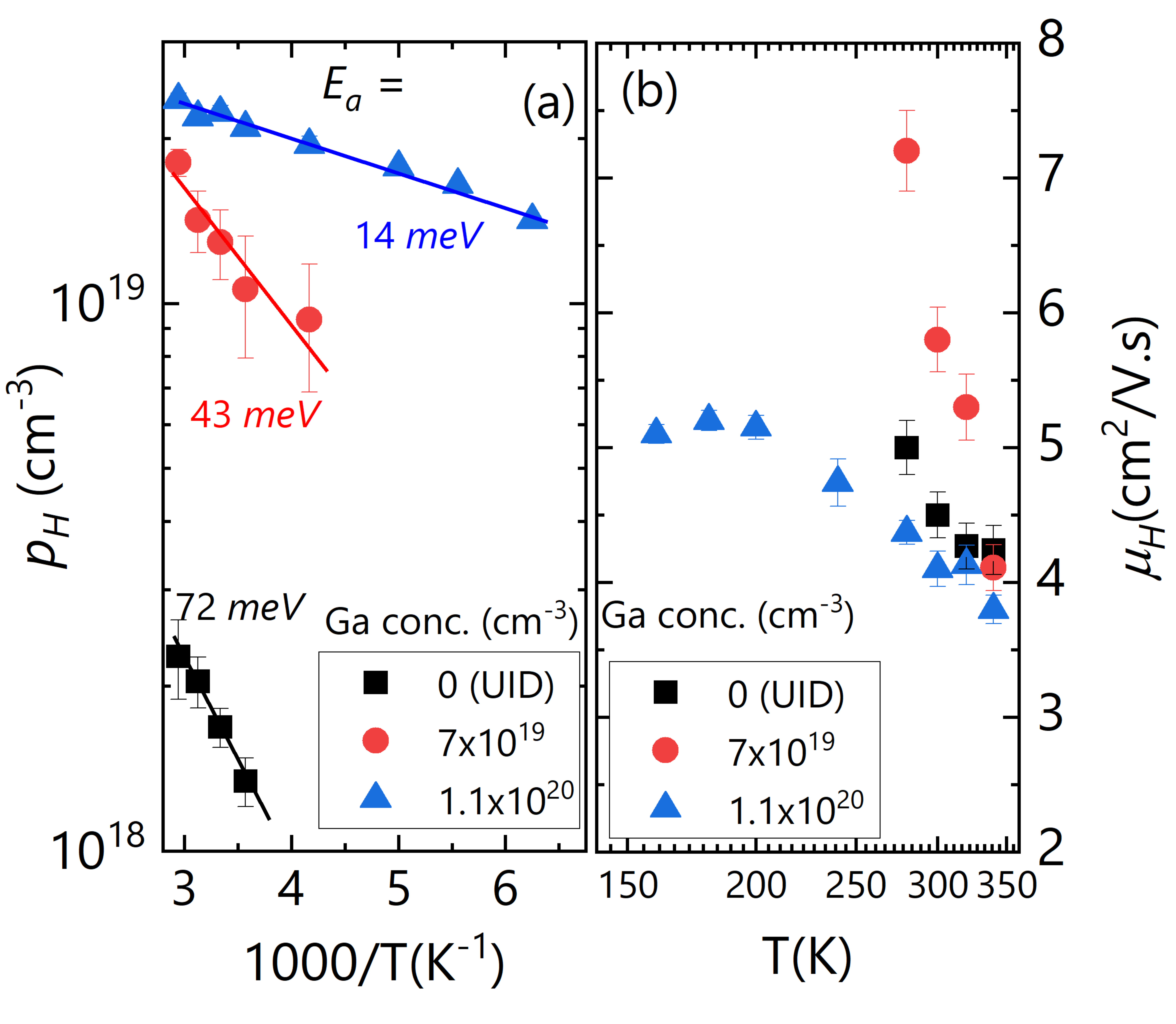}\centering
\caption{Hole transport properties of S-MBE grown SnO(001) thin films determined by van-der-Pauw-Hall measurements at temperatures T in the range of 150 to 340K. (a) Hole concentration for UID SnO and different Ga-doping levels as a function of temperature. The carrier concentration is shown in an Arrhenius plot with fits for the approximate acceptor activation energies, Ea. (b) Corresponding Hall mobilities of UID and Ga doped layers.\label{var_temp_hall}}
\end{figure}

The approximate acceptor activation energies are determined from the temperature-dependent Hall measurements as shown in Figure 4 for S-MBE grown samples. The Arrhenius plot of the hole concentration for UID SnO and SnO with different Ga-doping levels indicates an apparent activation energy of 74 meV for UID SnO grown via S-MBE. The activation energy decreases from 43 meV to 14 meV with increasing Ga concentrations from $7.0\times10^{19}$~cm$^{-3}$ to $1.1\times10^{20}$~cm$^{-3}$, indicating a likely impurity band formation near the valence band with increasing acceptor doping. To estimate the UID acceptor concentration, the temperature dependent Hall data for the UID sample is extrapolated to $T = \infty$ ($1/T=0$) indicating a UID acceptor concentration of $1.0\times10^{19}$~cm$^{-3}$.\cite{wbtj09}

The temperature dependent hole mobility in Figure 4(b) shows a strongly increasing mobility with decreasing temperature for UID SnO and Ga-doped SnO with acceptor concentrations of $7.0\times10^{19}$~cm$^{-3}$, indicating dominant phonon scattering. For samples with higher Ga doping of $1.1\times10^{20}$~cm$^{-3}$, the influence of scattering due to ionized donors, which is dominant at low temperature, led to the decrease in the mobility at low temperatures.

To explain the acceptor(donor) doping of SnO(001) with Ga(La), it is proposed that the ionic radii of the elements in different charge states promote the solubility and substitution of Ga$^{1+}$ (La$^{3+}$) in the Sn$^{2+}$ lattice leading to the observed acceptor(donor) doping behavior\cite{gfbrlpe17}. While Ga prefers the more stable 3+ state, 1+ states have been shown to be obtainable during molecular beam epitaxy by the formation of Ga$_2$O on the growth front or in the effusion cell.\cite{hcbo21,Vogt,vhactpmpbhbjxmmsld21} Interestingly, as shown in Table S1 (\textcolor{blue}{Supplementary material}) the ionic radius of Ga$^{1+}$ (113~pm) is close to that of Sn$^{2+}$ (118~pm), while Ga$^{3+}$ has a drastically different ionic radius of 62 pm. From a structural perspective, the similarity in the ionic radii of Ga$^{1+}$ and Sn$^{2+}$ could suggest that the substitutional incorporation of Ga$^{1+}$ will be favored over Ga$^{3+}$, thus promoting acceptor doping of SnO(001) with Ga$_\text{Sn}$. Also, La$^{3+}$ with an ionic radius of 103 pm should be structurally favored over La$^{1+}$ with ionic radius of 139 pm, leading to the observed compensating donor doping in SnO:La. 

To further understand the behavior of these and other dopants, their ab-initio-calculated formation energies and charge transition levels for Sn-rich and O-rich conditions are shown in Figure 5. The results suggest that La$_\text{Sn}$ act as a shallow donor in SnO. However experimental data indicate that semi-insulating rather than theoretically expected $n$-type conductive films are obtained as the La concentration exceeds the UID acceptor concentration. This observation suggests that the incorporated La$_\text{Sn}$ is being compensated by an increasing acceptor concentration relative to UID samples, such as V$_\text{Sn}$ or related complexes. This could be a result of the Fermi level position during growth, where higher concentrations of La$_\text{Sn}$ donors shift the Fermi level deeper into the band gap, facilitating increased incorporation of native acceptor compensators. Fig. 5 also indicates that Ga$_\text{Sn}$ and In$_\text{Sn}$ act most favorably as acceptors. Relative to the valence band maximum (VBM), the calculated $\epsilon$(0/\textendash) transition levels are 0.25 eV for Ga$_\text{Sn}$ and 0.13 eV for In$_\text{Sn}$, suggesting In$_{Sn}$ to be a more effective acceptor than Ga$_\text{Sn}$. While uncertainties on the order of 0.1~eV can be expected for the calculations, Ga$_\text{Sn}$ and In$_\text{Sn}$ acceptors are likely to exhibit incomplete ionization at room temperature.
An additional $\epsilon$(+/0) donor state exists approximately 0.02~eV below the VBM for Ga$_\text{Sn}$ and 0.04~eV below the VBM for In$_\text{Sn}$, which suggests that these dopants may be effectively pinning the Fermi level in the vicinity of the VBM through self-compensation. The results are consistent with the experimental observation of a saturating hole concentration with increasing Ga doping owing to the fact that it exhibits a localized electronic character. The stabilization of 1+ oxidation states for the Ga dopant is also in contrast to the other group III dopants like Al and B, which we theoretically find to preferentially act as shallow donors. Further measurements are needed to clarify the dominant electronic contributions and ionization energies of these dopants, as well as the role of compensator species that may be simultaneously incorporated to counteract the intended doping.

\begin{figure}
\includegraphics[width=8cm]{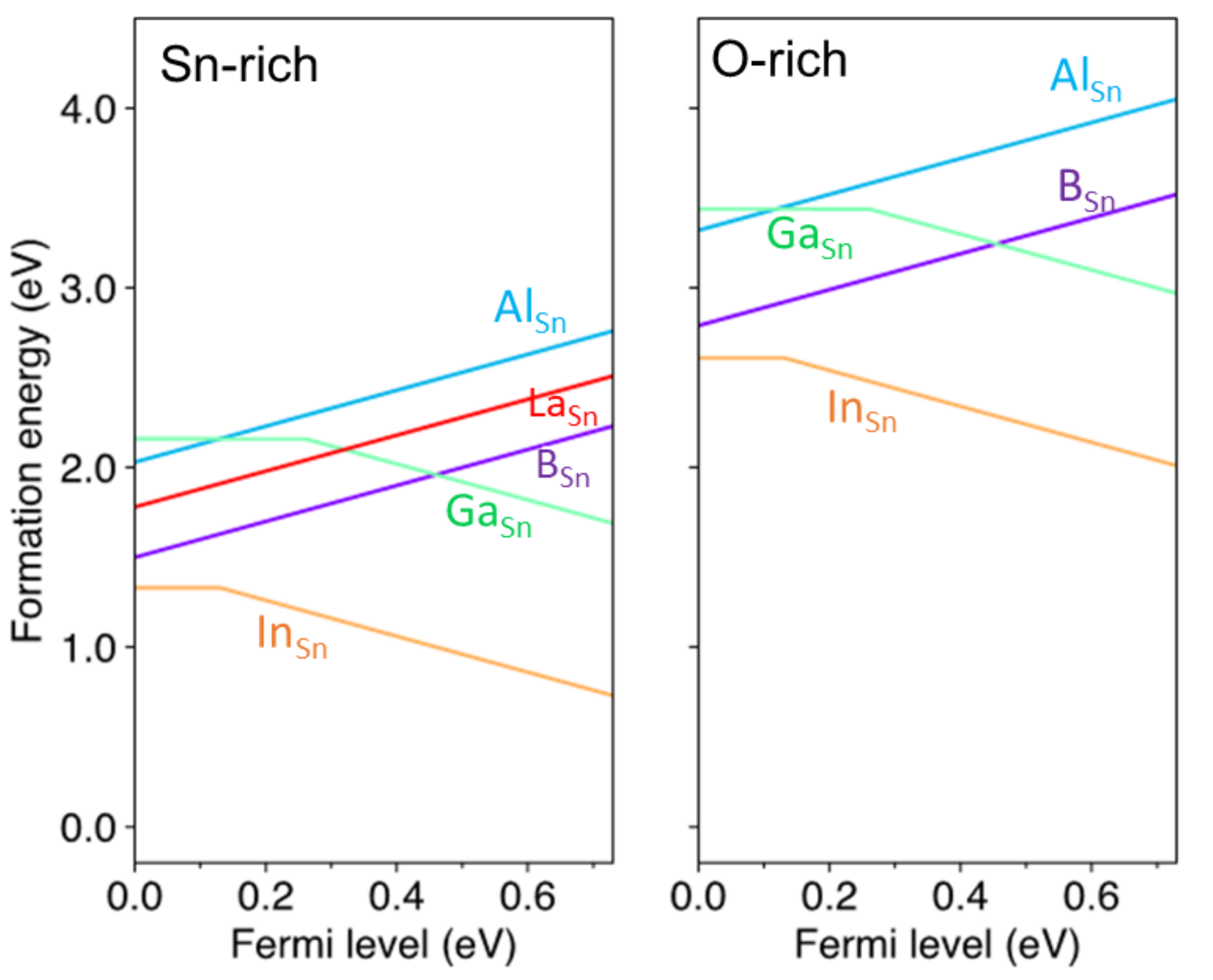}\centering
\caption{Ab-initio-calculated formation energies of In, Ga, La, Al and B substitutional dopants in SnO for Sn-rich and O-rich conditions.\label{fig:dft}}
\end{figure}

In conclusion, we have shown in this study that SnO (001) can be controllably doped using La and Ga atoms. X-ray diffraction data show the solubility limits of Ga and La in SnO are between $6.4$--$12\times10^{20}$~cm$^{-3}$ and $1.4$--$9.2\times10^{20}$~cm$^{-3}$ respectively. The hole concentration $p_H$ increases to $5.0\times10^{19}$~cm$^{-3}$ and $\rho$ decreases to 0.063 $\Omega$cm for PA-MBE-grown, Ga-acceptor-doped SnO(001) films. While SnO (001) films grown via S-MBE show a lower unintentional $p_H=1.2\times10^{18}$~cm$^{-3}$ and, $\rho=4.9$~$\Omega$cm  and $p_H$ increases up to $3.0\times10^{19}$~cm$^{-3}$ and $\rho$ decreases to 0.07 $\Omega$cm with Ga doping. In contrast, thin films doped with higher concentrations of La show a reduction in $p_H$ and a remarkable increase of $\rho$ up to 580 $\Omega$cm  without transition to $n$-type conductivity. Computational results identify that Ga and In preferentially act as deep acceptors with modest ionization energies in SnO, while La, Al, and B act as donors. Our results reveal that $p$-type conductivity in SnO can be controlled by intentional Ga- and La-doping over several orders of magnitude, whereas the successful $n$-type conduction of SnO remains challenging. 

By extending the boundary of conductivity in SnO across highly conductive SnO by Ga doping and semi-insulating SnO by La doping, we demonstrated material that can be applied in a wide variety of devices: SnO-based  $p$-channel thin film transistors can thus be optimized and even tuned from normally-on (high hole concentration) to normally-off (semi-insulating SnO)\cite{Hung2018, Zhang2022}.  Highly conductive Ga-doped SnO can be applied in $p$-type low resistance transparent contacts as well as promote the formation of the depletion zones inside (ultra-)wide bandgap $n$-type oxides in heterojunction $pn$-diodes and field effect transistor devices.\cite{tekupcagtbw22}
Last but not least, La-doping of SnO can lead to highly sensitive thin film conductometric gas sensors by reducing the gas-insensitive bulk conductivity of the film \cite{rombach_role_2016} which is electrically in parallel to the gas-sensitive, conductive surface.\cite{barsan_modeling_2010}

See the \textcolor{blue}{Supplementary material} for details of the EDX measurement and AFM images of UID and doped SnO(001) thin films.

The authors thank H.-P. Sch\"onherr and C. Hermann for MBE support, G. Hoffmann and A. Ardenghi for useful discussions and D. Dinh for critically reading the manuscript. This work was performed in the framework of GraFOx, a Leibniz-ScienceCampus partially funded by the Leibniz association. The work by J.B.V was performed under the auspices of the U.S. DOE by Lawrence Livermore National Laboratory under contract DE-AC52-07NA27344.

\bibliography{DopedSnO}

\end{document}